\documentclass[twocolumn,showpacs,preprintnumbers,amsmath,amssymb]{revtex4}

\usepackage{graphicx}
\usepackage{dcolumn}
\usepackage{bm}

\begin{document}


\title
{Spin-Polarized Transport through Double Quantum Dots
}

\author{
Yoichi {\sc Tanaka}  and  Norio {\sc Kawakami}
}

\affiliation{%
Department of Applied Physics, Osaka University, Suita, Osaka 565-0871
}%

\date{\today}

\begin{abstract}
We investigate spin-polarized transport phenomena through 
double quantum dots coupled 
to ferromagnetic leads in series. By means of the slave-boson mean-field 
approximation, we calculate the conductance 
in the Kondo regime for two different configurations of the leads: 
spin-polarization of two ferromagnetic leads is
 parallel or anti-parallel. It is found that transport 
shows some remarkable properties depending on the
tunneling strength between two dots. These properties are
explained in terms of the Kondo resonances
 in the local density of states. 
\end{abstract}


\maketitle

\section{Introduction}

Since the Kondo effect was observed in a quantum dot system\cite{Gold,Cronen}, 
electron correlation effects in such nanoscale systems have 
attracted much attention.
It was uncovered that the Kondo effect 
gives rise to remarkable properties in electron transport, which are not 
expected for free electron systems\cite{Glaz,Ng,MW,Kawa,Ogu,Sakai}.
For example, it makes tunneling transport possible at low temperatures
even if the system is in  the Coulomb blockade regime.
Since the Kondo effect originates from many body effects 
due to internal spin degrees of freedom,  transport properties are
quite sensitive to applied magnetic fields\cite{Gold,Cronen}.
This fact has naturally stimulated intensive works on 
spin-dependent transport phenomena
for a quantum dot connected to ferromagnetic leads
\cite{p1,p2,Zhang,Utumi,JiMa,Lopez,BinDon,Choi,Koing},
giving  interesting examples in  a
 rapidly developing field of magneto-electronics
or spin-electronics\cite{Prinz,Wolf,Fert,Julli,Slon}.

When several quantum dots are connected to each other,
they provide remarkable  phenomena due to
the interplay of electron correlations, interference effects, etc, which 
depend on how the dots are arranged: e.g. double 
quantum dots coupled in series or
parallel\cite{Lang,George,Rosa,Eto,Eto2,Izumi,Jeong,Chen},
T-shaped double dots\cite{Tae,Kikoin,Taka}, etc. In particular,
 the Kondo effect in  double quantum dot systems
 has been observed experimentally\cite{Jeong,Chen},
opening prospects for exploring further exotic phenomena 
due to electron correlations in nanoscale systems.

In this paper, we study spin-dependent transport through
a system of double quantum dots
coupled to two ferromagnetic (FM) leads in series.
Since the Kondo effect is sensitive to
 the presence of the FM leads, a variety 
of remarkable phenomena are expected to
 appear in transport at low temperatures.
By applying the slave-boson mean-field approximation to
the highly correlated Anderson model, we 
investigate  the equilibrium and nonequilibrium 
transport with the use of Keldysh Green function techniques.
We show that the strength of the inter-dot tunneling is
an important key parameter to  control
 transport phenomena via the modified Kondo resonances.
We further discuss how the modified Kondo resonances affect
transport properties for up- or down-spin electrons 
in the parallel and anti-parallel configurations
of the FM leads.

The organization of this paper is as follows. In the 
next section, we describe
the model and the method. Then in \S 3,
 we present the results for two different configurations: 
spin-polarization of two FM leads is
 parallel or anti-parallel. A brief summary is given in the 
last section.

\section{Model and Method}

We consider spin-dependent transport properties of a system with double 
quantum dots coupled in series.
\begin{figure}[h]
\begin{center}
\includegraphics[scale=0.38]{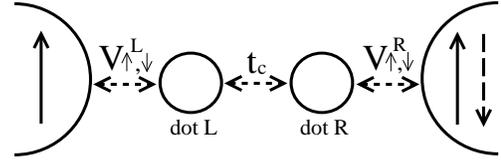}
\end{center}
\caption{Schematic picture of the double dot system coupled to
the FM leads:
$t_c$ is the interdot coupling and $V_{\alpha\sigma} 
(\alpha =L,R, \sigma=\uparrow, \downarrow)$ 
represents the dot-lead tunneling 
for an electron with spin $\sigma $.
}
\label{model}
\end{figure}
In Fig.\ref{model} we schematically show our double-dot
system. It is assumed that the left and right leads are 
spin-polarized, where the
 parallel (P) and anti-parallel (AP) configurations are considered.
In the following discussions,  the intradot  Coulomb 
interaction $U$ is 
 sufficiently large, so that double occupancy at each dot is 
forbidden, which allows us to use
 slave-boson representation\cite{AC,Cole} of correlated
electrons in the dots.
In terms of this representation, we can model the system in
Fig.\ref{model} with a $N$(=2) fold degenerate Anderson Hamiltonian
including an additional term for interdot coupling,
\begin{eqnarray}
H &=& \sum_{k_\alpha ,\sigma}\varepsilon _{k_\alpha\sigma}
c_{k_\alpha\sigma}^\dag c_{k_\alpha\sigma}^{}
 +\sum_{\alpha ,\sigma}\varepsilon _{\alpha\sigma}
f_{\alpha\sigma}^\dag f_{\alpha\sigma}^{}
\nonumber\\
 &+& \frac{t_c}{N}\sum_{\sigma}(f_{L\sigma}^\dag b_L^{} 
b_R^\dag f_{R\sigma}^{}+f_{R\sigma}^\dag b_R^{} b_L^\dag
 f_{L\sigma}^{})
\nonumber\\
& +& \frac{1}{\sqrt{N}}\sum_{k_\alpha ,\sigma} V_{\alpha\sigma}
(c_{k_\alpha\sigma}^\dag b_\alpha^\dag f_{\alpha\sigma}^{}+
 f_{\alpha\sigma}^\dag b_\alpha^{} c_{k_\alpha\sigma}^{})
\nonumber\\
 &+& \sum_{\alpha}\lambda_\alpha (\sum_{\sigma}f_{\alpha\sigma}^\dag 
f_{\alpha\sigma}^{}+b_\alpha^\dag b_\alpha^{} -1),\,
\label{Hami}
\end{eqnarray}
where $c_{k_\alpha\sigma}^\dag$($c_{k_\alpha\sigma}$) is
the creation (annihilation) operator for an electron with spin $\sigma$ 
in the  lead $\alpha$ ($\alpha=L, R$).
According to the slave-boson representation, the creation (annihilation)
operator of electrons in the left or right dot, 
$d_{\alpha\sigma}^\dag$($d_{\alpha\sigma}$), is replaced by
$d_{\alpha\sigma}^\dag\to f_{\alpha\sigma}^\dag b_{\alpha}^{}$
($d_{\alpha\sigma}\to f_{\alpha\sigma}^{}b_{\alpha}^\dag $),
where $b_{\alpha}$ ($f_{\alpha\sigma}$) is the slave-boson
(pseudo-fermion) annihilation operator for an empty state 
(singly occupied state).
The last term with the Lagrange multiplier
$\lambda_\alpha$ is introduced so as to 
incorporate the constraint
imposed on the slave particles,
$\sum_{\sigma=\uparrow, \downarrow}
 {f_{\alpha\sigma }^\dag f_{\alpha\sigma }^{}}+b_\alpha^\dag b_\alpha^{}=1$.
For simplicity, we consider the symmetric dots with
$\varepsilon _{L\sigma}=\varepsilon _{R\sigma}=\varepsilon _{0}$.
The mixing term $V_{\alpha\sigma}$ in  \eqref{Hami} leads 
to the linewidth function
\begin{equation}
\Gamma_{\sigma}^\alpha(\varepsilon )=\pi \sum_{k_{\alpha},\sigma}^{}
|V_{\alpha ,\sigma}|^2\delta(\varepsilon -\varepsilon_{k_{\alpha}\sigma}),
\label{ganma}
\end{equation}
which  is reduced to an energy-independent constant
 $\Gamma_{\sigma}^\alpha$ in the wide band limit.
 Here, we introduce  the effective 
spin-polarization strength \textit{p}, following the 
 definition given in the literature:~\cite{p1,Zhang,Utumi,JiMa,BinDon}
\begin{equation}
p=\frac{\Gamma_{\sigma}^\alpha-\Gamma_{\bar{\sigma}}^\alpha}{\Gamma_{\sigma}^\alpha+\Gamma_{\bar{\sigma}}^\alpha}\,
 \,\, (0\le p\le 1)\,.
\label{p}
\end{equation}
This quantity is convenient to represent how large the 
strength of the spin polarization is.
From the definition of \textit{p}, $\Gamma_{\sigma}^\alpha$ is 
expressed in the case of the P and AP
configurations, 
\begin{eqnarray}
{\rm P}:\Gamma_{\uparrow}^L=\Gamma_{\uparrow}^R=(1+p)\Gamma_0
\,,\,\Gamma_{\downarrow}^L=\Gamma_{\downarrow}^R=(1-p)\Gamma_0
\label{resonance1}
\\
{\rm AP}:\Gamma_{\uparrow}^L=\Gamma_{\downarrow}^R=(1+p)\Gamma_0
\,,\,\Gamma_{\downarrow}^L=\Gamma_{\uparrow}^R=(1-p)\Gamma_0,
\label{resonance2}
\end{eqnarray}
where $\Gamma_0$ is the value at $p=0$.

To analyze our model, we use the slave-boson mean-field 
approximation~\cite{AC,Cole,Rosa,Lang,Eto}, in which boson
fields are approximated by their mean values,
 $b_\alpha(t)/\sqrt{N}\to \langle b_\alpha(t)\rangle /\sqrt{N}
=\tilde{b}_\alpha$. 
By introducing the renormalized quantities
$\tilde{V}_{\alpha\sigma}=V_{\alpha\sigma}\tilde{b}_\alpha$, 
$\tilde{t}_c=t_c\tilde{b}_L\tilde{b}_R$,
and $\tilde{\varepsilon}_{\alpha}=\varepsilon_{0}+\lambda_\alpha$,
the total Hamiltonian \eqref{Hami} is now replaced by
\begin{eqnarray}
\tilde{H}&=& \sum_{k_\alpha ,\sigma}\varepsilon _{k_\alpha\sigma}
c_{k_\alpha\sigma}^\dag c_{k_\alpha\sigma}^{}
 +\sum_{\alpha ,\sigma}\tilde{\varepsilon}_{\alpha}
f_{\alpha\sigma}^\dag f_{\alpha\sigma}^{}
\nonumber\\
 &+& \tilde{t_c}\sum_{\sigma}(f_{L\sigma}^\dag f_{R\sigma}^{}+
f_{R\sigma}^\dag f_{L\sigma}^{})
\nonumber\\
&+& \sum_{k_\alpha ,\sigma}\tilde{V}_{\alpha\sigma}
(c_{k_\alpha\sigma}^\dag f_{\alpha\sigma}^{}
 +f_{\alpha\sigma}^\dag c_{k_\alpha\sigma}^{})
\nonumber\\
&+&\sum_{\alpha}\lambda_\alpha (N\tilde{b}_\alpha^2-1).
\label{tilHami}
\end{eqnarray}
To determine the mean-field values of 
($\tilde{b}_L,\tilde{b}_R,\lambda_L,\lambda_R$),
we employ the equation of motion method
 with the non-equilibrium Keldysh  Green functions.  
This gives the 
following set of self-consistent equations,
\begin{align}
&\tilde{b}_{L(R)}^2-i\sum_\sigma\int\frac{d\varepsilon }{4\pi}
G_{L,L(R,R)\sigma}^<(\varepsilon )=\frac{1}{2},\\
&(\tilde{\varepsilon}_{L(R)}-\varepsilon_0)\tilde{b}_{L(R)}^2
\nonumber\\
&\qquad\quad -i\sum_\sigma\int\frac{d\varepsilon }{4\pi}
(\varepsilon -\tilde{\varepsilon}_{L(R)})\,
G_{L,L(R,R)\sigma}^<(\varepsilon )=0,
\label{self}
\end{align}
where $G_{L,L(R,R)\sigma}^<(\varepsilon )$ is the Fourier 
transform of the Keldysh Green function
$G_{\alpha,\alpha '\sigma}^<(t-t')\equiv 
i\langle f_{\alpha '\sigma}^\dag (t')f_{\alpha\sigma}^{}(t)\rangle$.
The first equation represents the constraint imposed on
the slave particles, while the second one is obtained from 
the stationary condition that the boson field is
 time-independent at the mean-field level.
From the equation of motion  of the operator 
$f_{\alpha\sigma}$ ~\cite{Keldysh,Lang}, we 
have the explicit form of the Green function,
\begin{widetext}
\begin{align}
G_{L,L(R,R)\sigma}^<(\varepsilon )=\frac
{2i\left\{f_{L(R)}(\varepsilon )\tilde{\Gamma}_{\sigma}^{L(R)}
\left[(\varepsilon -\tilde{\varepsilon}_{R(L)})^2
+(\tilde{\Gamma}_{\sigma}^{R(L)})^2\right]
+f_{R(L)}(\varepsilon )\tilde{\Gamma}_{\sigma}^{R(L)}{\tilde{t_c}}^2\right\}}
{\left|(\varepsilon-\tilde{\varepsilon}_{L}+i\tilde{\Gamma}_{\sigma}^L)
(\varepsilon-\tilde{\varepsilon}_{R}+i\tilde{\Gamma}_{\sigma}^R)
-{\tilde{t_c}}^2\right|^2}
\label{lessG}
\end{align}
\end{widetext}
 with
$\tilde{\Gamma}_{\sigma}^\alpha=
\tilde{b}_{\alpha}^2\Gamma_{\sigma}^\alpha \,(\alpha =L,R)$.
Note here 
that we have derived the above formulae
by applying the analytic  continuation rules to the
equation of motion of the time-ordered Green functions along a 
complex contour.

By using the renormalized parameters determined self-consistently, 
we obtain the current $I$ through the 
two dots,~\cite{Lang,Keldysh,Eto,Eto2}
\begin{eqnarray}
I=\frac{2e\Gamma_0}{h}\sum_{\sigma}
\int d\varepsilon \,(f_L(\varepsilon )
-f_R(\varepsilon ))\,T_\sigma(\varepsilon )
\label{I}
\end{eqnarray}
with the Fermi function $f_\alpha(\varepsilon)$,
where the transmission probability $T_\sigma(\varepsilon )$ 
for spin $\sigma $ is
\begin{eqnarray}
T_\sigma (\varepsilon )=\frac
{2{\tilde{t_c}}^2\tilde{\Gamma}_{\sigma}^L\tilde{\Gamma}_{\sigma}^R}
{\left|(\varepsilon-\tilde{\varepsilon}_{L}+i\tilde{\Gamma}_{\sigma}^L)
(\varepsilon-\tilde{\varepsilon}_{R}+i\tilde{\Gamma}_{\sigma}^R)
-{\tilde{t_c}}^2\right|^2}\,\,.
\label{TP}
\end{eqnarray}

We can thus compute the differential conductance $G=dI/dV$ from
the current $I$ in \eqref{I}.
Moreover, we obtain the density of states (DOS) 
for the dot $\alpha$ ($\alpha=L$ or $R$) as 
\begin{eqnarray}
\rho_{\alpha,\sigma} (\varepsilon )=-\frac{1}{\pi} {\tilde{b}_{\alpha}}^2
\textrm{Im}G_{\alpha,\alpha\sigma}^{r} (\varepsilon ),
\label{DOS}
\end{eqnarray}
where $G_{\alpha,\alpha\sigma}^{r} (\varepsilon )$ is the Fourier 
transform of the Keldysh Green function
$G_{\alpha,\alpha\sigma}^r(t-t')\!\equiv\! 
-i\theta(t-t')\langle\{f_{\alpha\sigma}(t),f_{\alpha\sigma}^\dag(t')\}\rangle$, which is given by 
\begin{eqnarray}
G_{L,L(R,R)\sigma}^{r}\! (\varepsilon )\!=\!
\frac
{(\varepsilon-\tilde{\varepsilon}_{R(L)}+i\tilde{\Gamma}_{\sigma}^{R(L)})}
{(\varepsilon-\tilde{\varepsilon}_{L}+i\tilde{\Gamma}_{\sigma}^L)
(\varepsilon-\tilde{\varepsilon}_{R}+i\tilde{\Gamma}_{\sigma}^R)
-{\tilde{t_c}}^2}.
\label{rG}
\end{eqnarray}
This completes our formulation of the non-equilibrium
transport for the double-dot system with spin polarization.

\section{Numerical Results}
In this section, we discuss the results computed at 
absolute zero
for the P and AP configurations separately.
For simplicity, the band width of the leads is taken as  $D=60\Gamma_0$ 
and the bare level for two dots is fixed as
$\varepsilon _{0}=-3.5\Gamma_0$ (Kondo regime).
We will use $\Gamma_0 $ as the unit of the energy and 
choose the Fermi level at  $V=0$ as the origin of the energy.

\subsection{Parallel configuration}

Let us start our discussions with
 the parallel (P) configuration of the polarized leads.

\subsubsection{Equilibrium case $(V=0)$}

We first investigate transport properties of the equilibrium dot
system where the bias voltage $V$ is infinitesimally small.
\begin{figure}[h]
\includegraphics[scale=0.30]{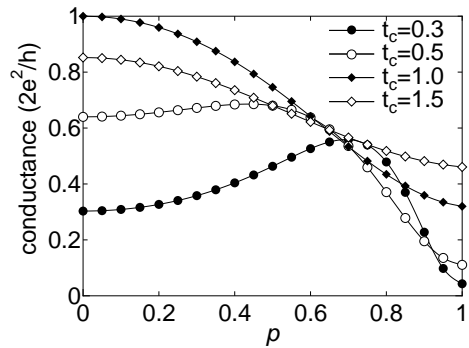}
\caption{Linear conductance $G_{V=0}=dI/dV|_{V=0}$ as a
function of the
spin-polarization strength $p$ for $t_c=0.3,0.5,1.0$ and $1.5$.
}
\label{Pp-g}
\end{figure}
In Fig. \ref{Pp-g}, we plot the linear conductance 
$G_{V=0}=dI/dV|_{V=0}$ as a function of the 
spin polarization strength $p$ for various interdot couplings.
For $t_c=0.3$, $G_{V=0}$ exhibits a maximum around $p=0.7$, whereas
 it decreases monotonically for $t_c=1.0$ and $1.5$.
To clearly observe  the characteristic $p$-dependence in the 
conductance,
let us consider the contribution of up- and down-spin electrons to $G_{V=0}$.
\begin{figure}[h]
\includegraphics[trim=0cm 0mm 0mm 0mm,scale=0.3]{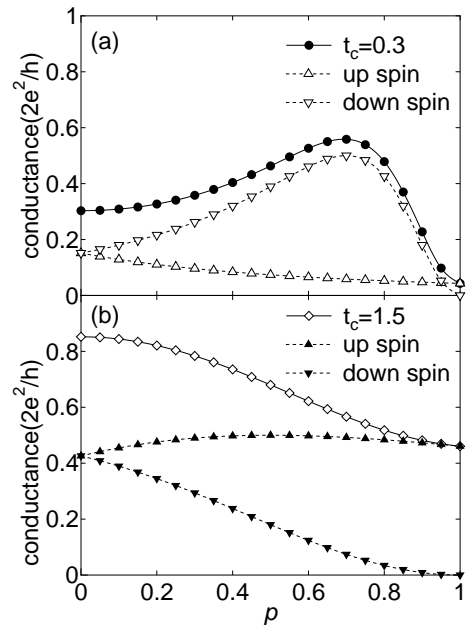}
\caption{
The contribution of up and down spin electrons to
the conductance for (a) $t_c=0.3$ and (b) $1.5$.
}
\label{Pp-gud}
\end{figure}
For small interdot coupling ($t_c=0.3$) in Fig. \ref{Pp-gud}(a), 
it is seen that down-spin (i.e. minority-spin) electrons contribute dominantly
 to $G_{V=0}$ at large $p$,  featuring
a maximum structure around $p=0.7$.
However, for larger interdot couplings (e.g. $t_c=1.5$) in
 Fig. \ref{Pp-gud}(b),  
$G_{V=0}$ gets dominated by up-spin (i.e. majority-spin) electrons 
as $p$ increases.
\begin{figure}[h]
\includegraphics[trim=-0cm 0mm 0mm 0mm,scale=0.38]{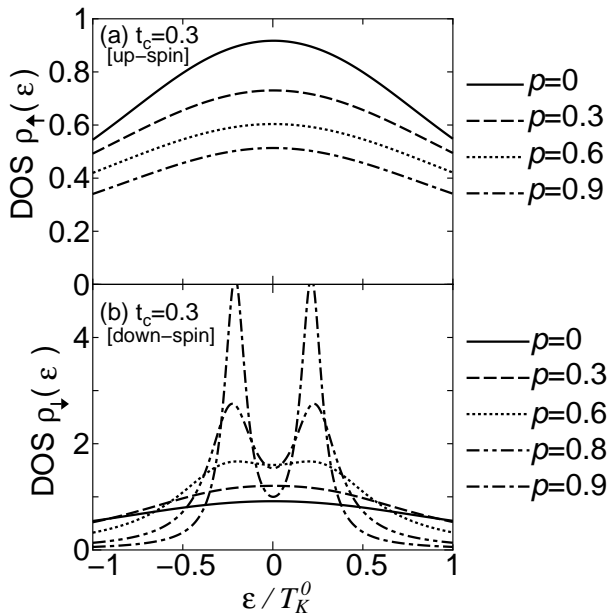}
\caption{ DOS for (a) up- and (b) down-spin electrons of the dots
 in the case of $t_c=0.3$.
The energy $\varepsilon$ is normalized by 
the Kondo temperature in the equilibrium case 
$T_K^0$ ($T_K^0\simeq 10^{-3}\Gamma_0$).
}
\label{dos03}
\end{figure}

\begin{figure}[h]
\includegraphics[trim=-0cm 0mm 0mm 0mm,scale=0.38]{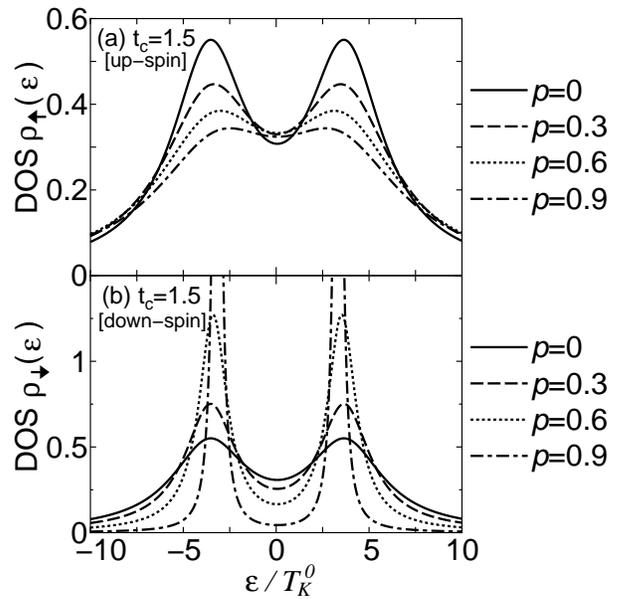}
\caption{
DOS of the dots for (a)  up- and (b) down-spin electrons 
in the case of $t_c=1.5$.}
\label{dos1.5}
\end{figure}

The origin of the above characteristic properties can be seen 
in the formation of the modified Kondo resonances in the presence of $p$.
Figures \ref{dos03} and \ref{dos1.5} show the local DOS  for up- and down-spin
 electrons of the dots around the Fermi energy $\varepsilon=0$.
In the equilibrium case with
$V=0$, we denote the DOS for the left and right 
dots as $\rho_{\sigma}(\varepsilon )=\rho_{L\sigma}(\varepsilon )
=\rho_{R\sigma}(\varepsilon )$.

In the absence of $p$, one can see the formation of the Kondo 
resonance around $\varepsilon=0$, which is renormalized about 
$10^{-3}$ smaller than the bare resonance. 
First we consider the case of $t_c=0.3$.
As the bare resonance width for up-spin 
electrons increases with the increase of $p$, according to
Eq.\eqref{resonance1},
the resulting Kondo resonance for $\rho_{\uparrow }(\varepsilon )$ 
 in Fig. \ref{dos03}(a) becomes broad,
so that  $\rho_{\uparrow }(0)$, which determines 
 $G_{V=0}$, decreases.
On the other hand, as seen from $\rho_{\downarrow }(\varepsilon )$ 
in Fig. \ref{dos03}(b),
the width of the Kondo resonance for down spin electrons becomes smaller with
 increasing $p$. Therefore, the corresponding
peak in $\rho_{\downarrow }(\varepsilon )$  splits into two sub-peaks
 when its width  gets comparable to  the renormalized 
interdot coupling $\tilde t_c$. Namely, the Kondo 
resonances in the left and right dots  are combined into
the bonding and antibonding  Kondo resonances, 
which are respectively located around
\begin{align}
\varepsilon_{\pm}=\frac{1}{2}\left\{(\tilde{\varepsilon}_{L}
+\tilde{\varepsilon}_{R})\pm
\sqrt{(\tilde{\varepsilon}_{L}-\tilde{\varepsilon}_{R})^2
+4\tilde{t_c}^2}\right\}.
\end{align}
Since $\tilde{\varepsilon}_{L}=\tilde{\varepsilon}_{R}$ 
at $V=0$,
the energy difference $2\tilde{t_c}(=\varepsilon_{+}- \varepsilon_{-})$ 
gives the  splitting of the Kondo resonances.  Thus,  
$\rho_{\downarrow}(0)$ increases first and then decreases with 
the increase of $p$. The above $p$-dependence of the Kondo resonances 
naturally explains the characteristic behavior of 
 $G_{V=0}$ in Fig. \ref{Pp-gud}(a).

We next discuss the case of $t_c=1.5$.
As seen in Fig. \ref{dos1.5}, the large interdot coupling
causes the splitting of the Kondo resonances even at $p=0$, 
which is obscured (sharpened) for up spins (down spins) with 
the increase of $p$.
As a result, $\rho_{\uparrow}(0)$ little changes, while 
$\rho_{\downarrow}(0)$ monotonically decreases as $p$ increases.
This clarifies why 
 the contribution of up-spin electrons to $G_{V=0}$ is almost 
independent of $p$ whereas
that of down spin electrons decreases with increasing $p$ 
in Fig. \ref{Pp-gud}(b).

When we change the strength of tunneling $t_c$,
the system changes gradually between the above two typical cases.  


\subsubsection{Nonequilibrium case $(V\ne 0)$}

We next show the results in the nonequilibrium case. 
By using Eq.\eqref{I}, we calculate the current $I$ through the 
system as a function of $V$: we set
the chemical potentials $\mu_L=V/2$ and $\mu_R=-V/2$.
We wish to mention here that
Aguado et al.  studied  transport 
properties for normal leads with finite bias voltage.\cite{Lang}

\begin{figure}[h]
\begin{minipage}{.75\linewidth}
\includegraphics[trim=0cm 0mm 0mm 0mm, scale=0.36]{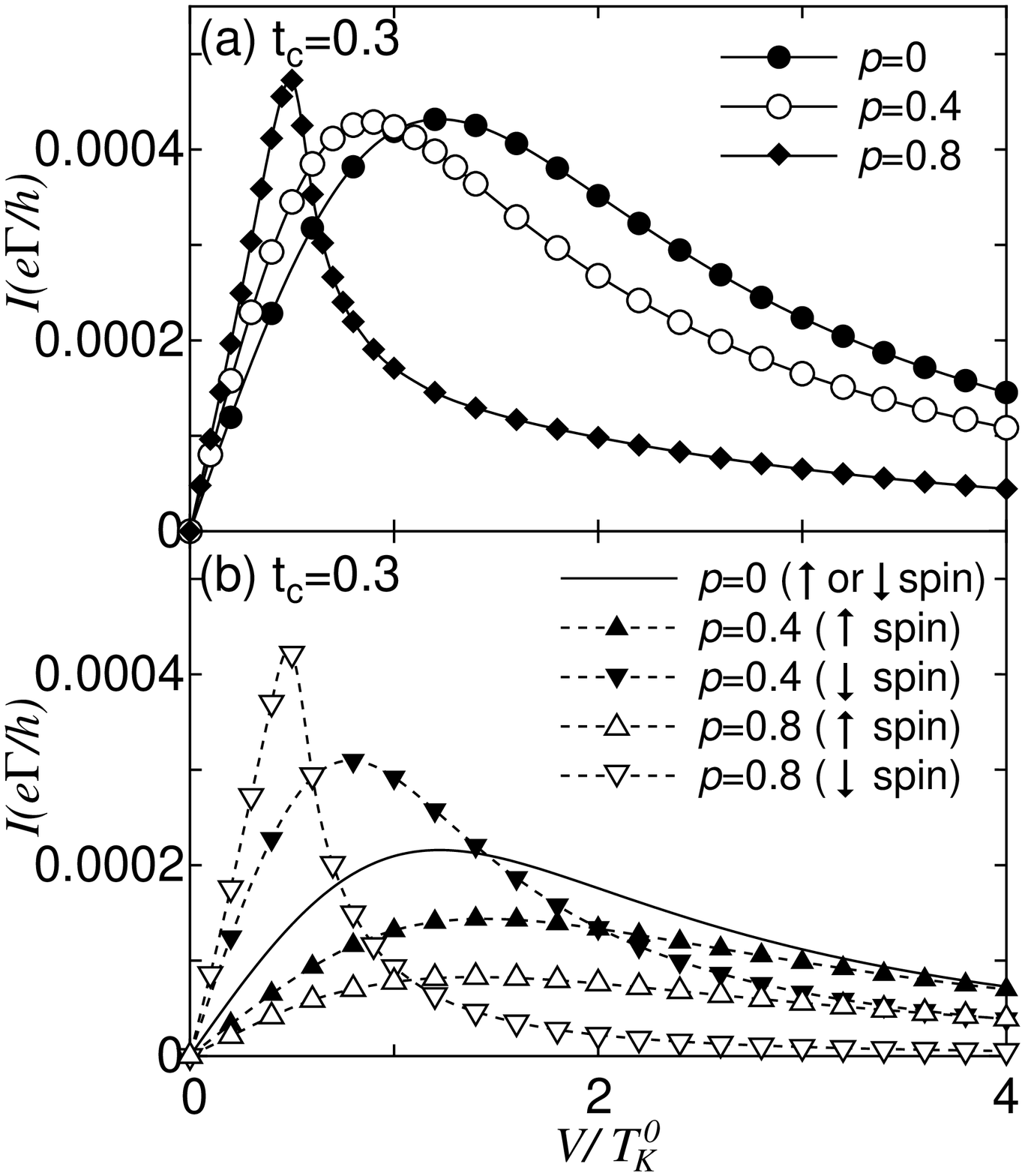}
\end{minipage}
\begin{minipage}{.75\linewidth}
\includegraphics[trim=0cm 0mm 0mm 0mm, scale=0.36]{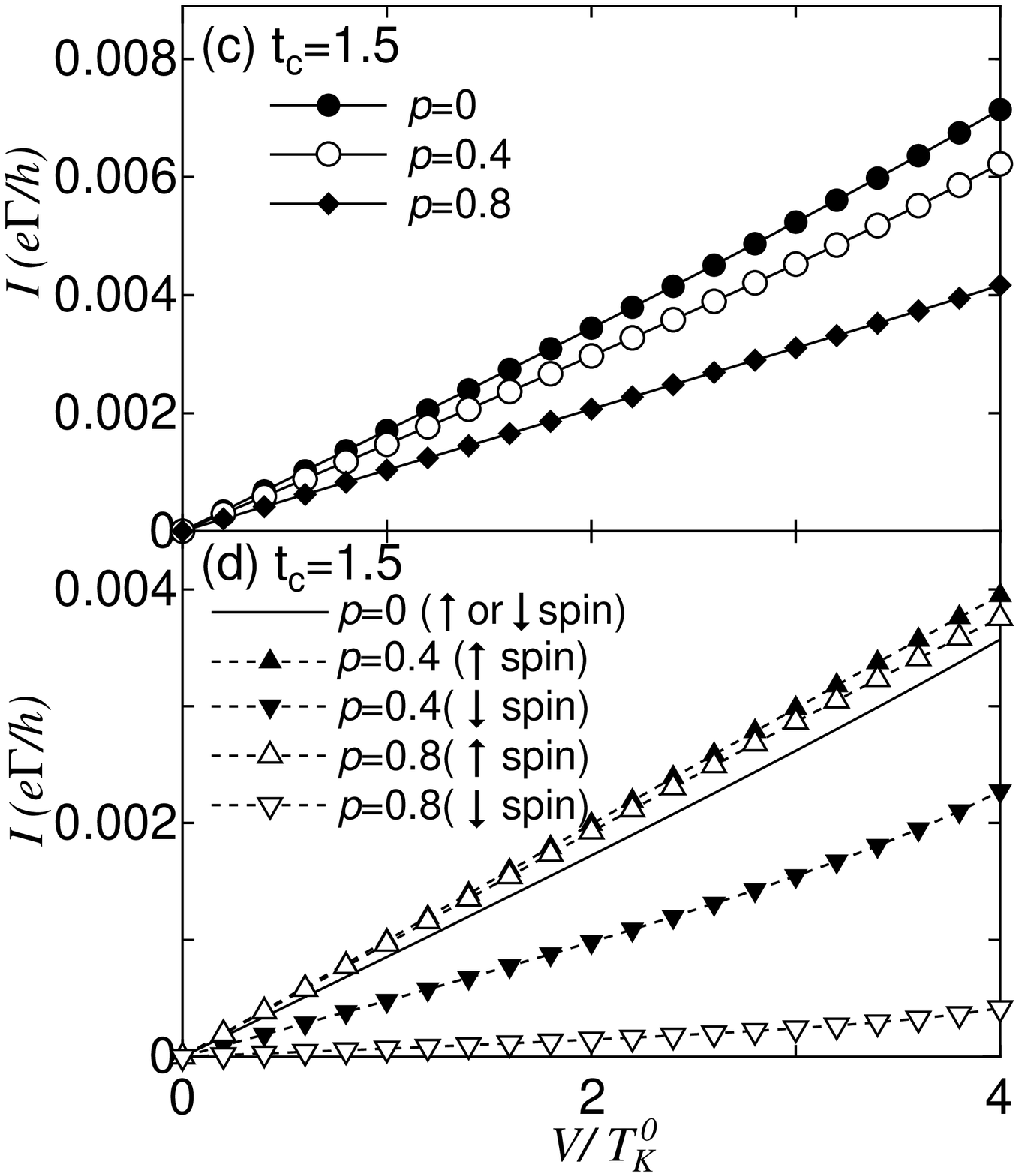}
\end{minipage}
\caption{
(a) $I$-$V$ characteristics  and (b) the contribution
of up- and down-spin electrons for
$t_c=0.3$.  Similar plots are shown in
(c) and (d) for $t_c=1.5$.
}
\label{Piv}
\end{figure}

Figure \ref{Piv} shows 
the $I$-$V$ characteristics as well as the contribution 
of up- and down-spin electrons. Note that the conductance
$G_{V=0}$ discussed above coincides with the gradient 
of the {\it I-V} curves at $V=0$ in Fig. \ref{Piv}.
For $t_c=0.3$, as shown in Fig. \ref{Piv}(a), 
the current once increases and then
decreases after making a peak structure
in its $V$ dependence. The peak structure becomes sharp
 with the increase of $p$, which is mainly caused by down-spin electrons,
as seen in Fig. \ref{Piv}(b).  We can thus see that for finite $p$
the current is dominated by down spin electrons for 
small $V$, while it is mainly contributed by
up-spin electrons for large $V$. In contrast, we encounter different 
behavior for $t_c=1.5$ in Figs. \ref{Piv}(c) and (d).
In the voltage regime shown in the figure,
the current $I$ is almost linear in $V$, and
is  dominated by up-spin electrons
as $p$ increases.

As in the equilibrium case, we interpret the above
 {\it I-V} characteristics in terms of  the modified Kondo resonances.
\begin{figure}[h]
\begin{minipage}{.50\linewidth}
\includegraphics[trim=-0.0cm 0cm 0cm 0mm,scale=0.35]{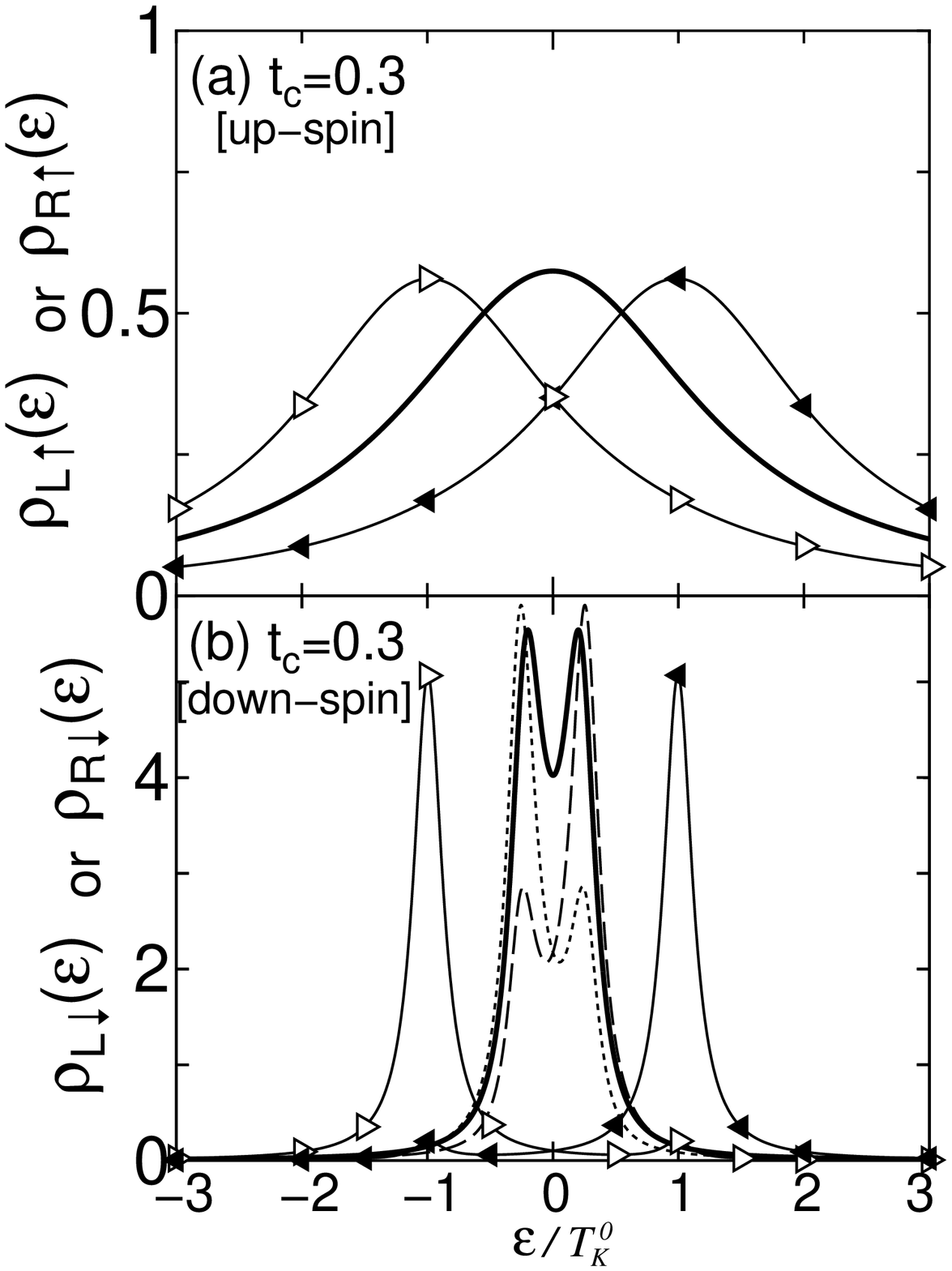}
\end{minipage}
\begin{minipage}{.45\linewidth}
\includegraphics[trim=-3.0cm 0cm 0mm 0mm,scale=0.36]{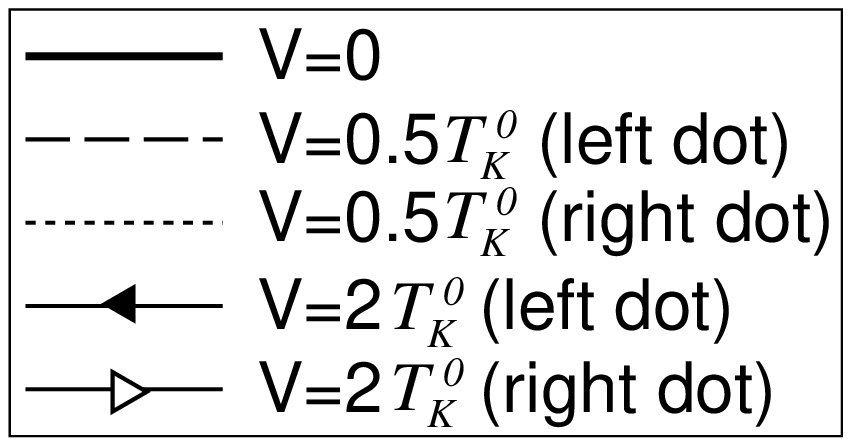}
\end{minipage}
\end{figure}
\begin{figure}[h]
\vspace{-0.3cm}
\begin{minipage}{.50\linewidth}
\includegraphics[trim=-0.4cm 0mm 0cm 0cm,scale=0.35]{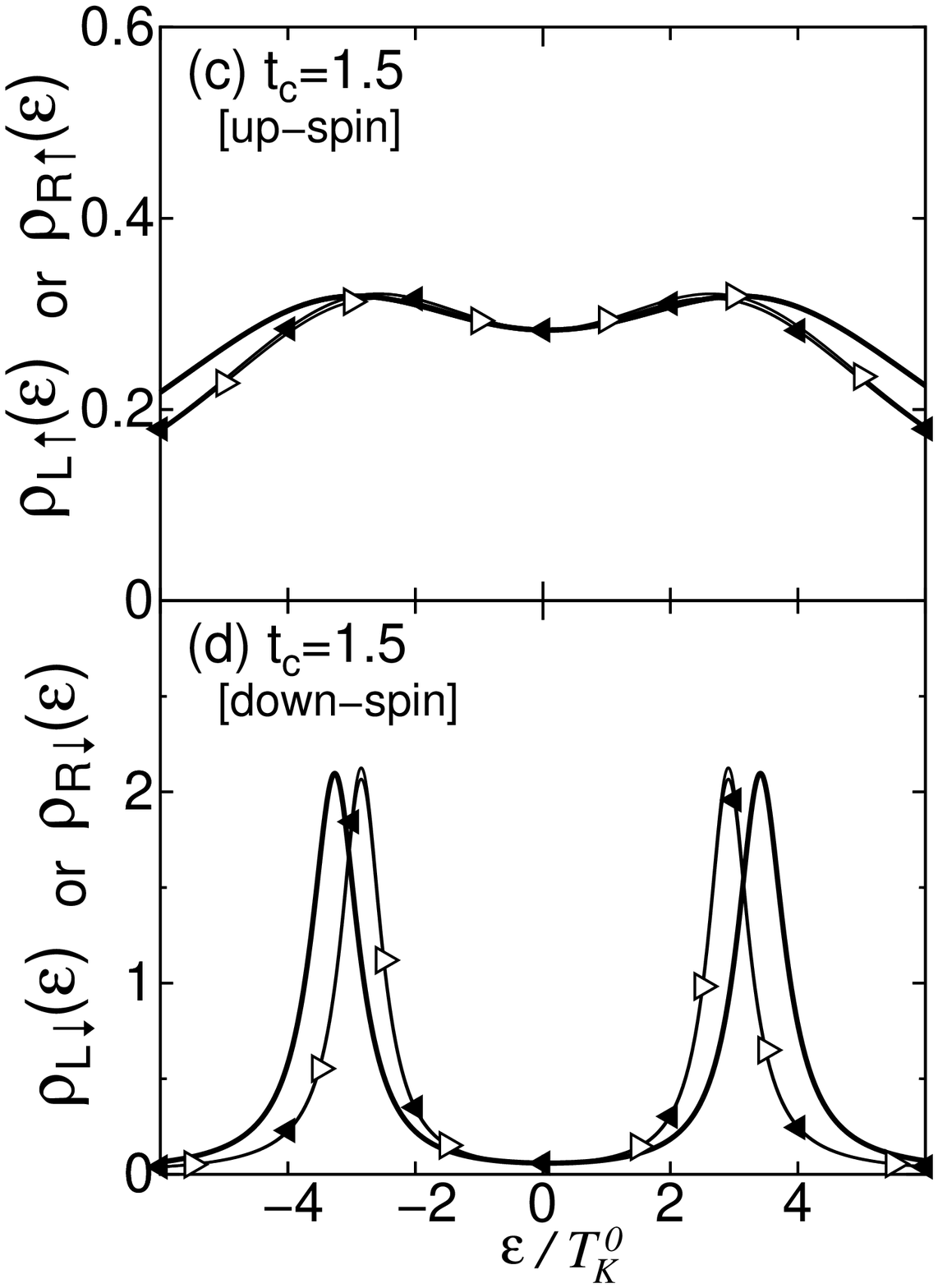}
\end{minipage}
\begin{minipage}{.45\linewidth}
\includegraphics[trim=-3.2cm 0mm 0mm 0cm,scale=0.36]{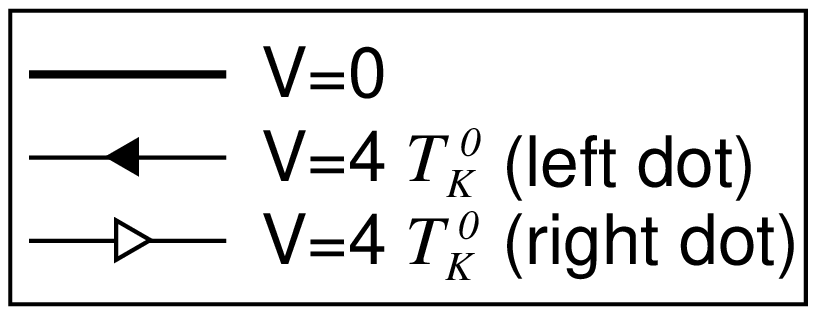}
\end{minipage}
\caption{
DOS for left and right dots at $p=0.8$ in the nonequilibrium
case: $t_c=0.3$  for (a) up- and (b) down-spin
electrons; $t_c=1.5$ for (c) up-  and (d)  down-spin electrons.
}
\label{Pdos}
\end{figure}
In Fig. \ref{Pdos} we show the Kondo resonances in the
nonequilibrium case 
by exploiting $t_c=0.3$ and $t_c=1.5$ with
a typical spin-polarization $p=0.8$.

Let us first look at  
 the Kondo resonance at $t_c=0.3$  in Figs. \ref{Pdos}(a) and (b). 
At $V=0$, the DOS for up-spin electrons shows a broad hump, while that 
for down-spin electrons has sharp peaks with small splitting
due to $t_c$.
With increasing $V$, the Kondo resonance of the left (right)
 dot is shifted, following change in the chemical potential
 of the left (right) lead. 
Roughly speaking,  the current $I$ depends
 on the overlap of these two Kondo resonances
 in the energy region between $\mu_L$ and $\mu_R$
($-V/2< \varepsilon < V/2$ in Fig. \ref{Pdos}).
By this reason, at small $V$, the current $I$ for up-spin 
electrons gradually increases with the increase of $V$
 and goes down reflecting the decrease in the 
overlap  of the Kondo resonances for larger $V$.
On the other hand, for down spins (Fig. \ref{Pdos}(b)),
the splitting of the resonances is enlarged
as $V$ increases, leading to the well-separated 
 peaks. Correspondingly, the current $I$
rises sharply for small $V$, and then goes down quickly as  
$V$ gets large. 

When the inter-dot coupling is large ($t_c=1.5$), we can 
see the splitting of the Kondo resonances both for up-spin 
and down-spin 
electrons at $V=0$, although they are well separated 
(Fig. \ref{Pdos}(d)) for down spins
but more or less obscured for up spins (Fig. \ref{Pdos}(c)).
  A remarkable point is that their shapes little change
 with increasing $V$,
because the coupling between the two dots is much larger
than the applied bias.  Note that
the DOS for up- and down-spin electrons is 
almost flat in the vicinity of the Fermi energy,
so that the current $I$ is approximately linear in the bias 
voltage and  mainly caused by up-spin electrons,
in accordance with the results of Figs. \ref{Piv}(c) and (d).
Here, we wish to mention what happens 
for larger voltage in the case of $t_c=1.5$.
We have checked that the system undergoes a first-order phase
transition when $V$ further increases beyond $V/T_K \sim 4$,
where the current drops 
discontinuously. This implies that a strongly
coupled molecule is formed between two dots, which is almost 
decoupled from two leads. Such a phenomenon was already pointed 
out  in the $p=0$ case.\cite{Lang} The detail of the first-order 
transition can be found there.

\subsection{Antiparallel configuration}

We now move to the system with the AP configuration
of the polarized leads.  We can continue discussions
similarly to  the above P case, so that we summarize 
the essential points briefly.

\subsubsection{Equilibrium case $(V=0)$}

\begin{figure}[h]
\includegraphics[trim=0cm 0mm -0cm 0mm,scale=0.41]{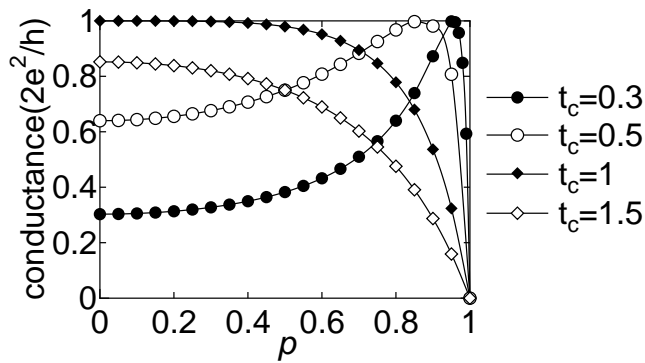}
\caption{Linear conductance $G_{V=0}$ as a function of 
the spin polarization strength $p$  for the AP configuration.
}
\label{APp-g}
\end{figure}
We start with  the equilibrium case.
In Fig. \ref{APp-g}, we plot the linear conductance $G_{V=0}$ 
as a function of  the
spin polarization strength $p$.
Note that the contribution of up- and down-spin electrons to 
the total conductance is same in the AP case. Nevertheless,
we observe the behavior somewhat analogous  to the 
case of the P configuration: for small $t_c$ the 
total conductance features  a peak structure in its 
$p$ dependence, while it  decreases monotonically for 
large $t_c$. On the other hand, for the fully polarized case with $p=1$,
the current vanishes in the present case in contrast  to the 
P configuration. 

\begin{figure}[h]
\begin{minipage}{.90\linewidth}
\includegraphics[trim=-0cm 0mm 0mm 0mm,scale=0.38]{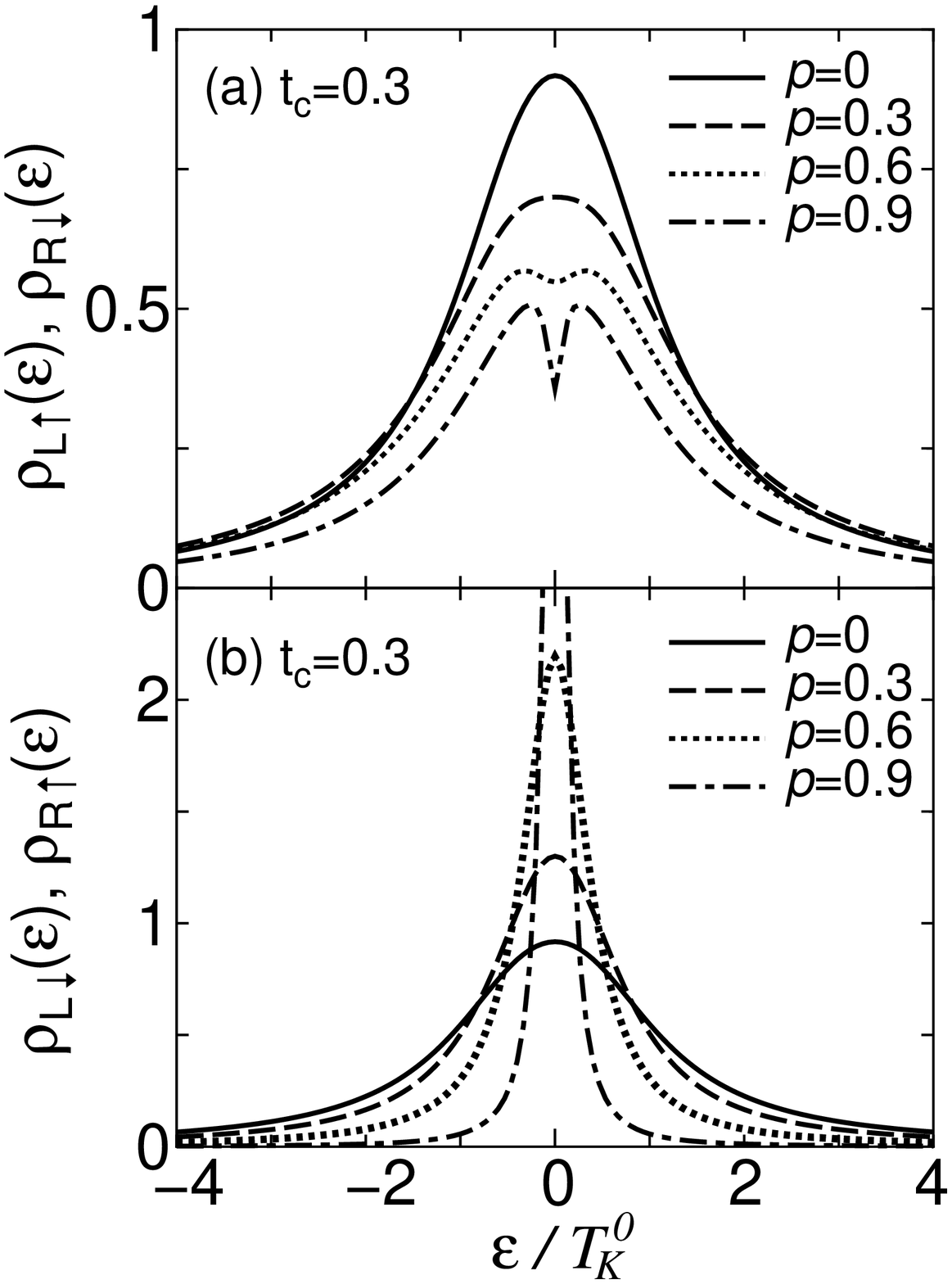}
\end{minipage}
\begin{minipage}{.90\linewidth}
\includegraphics[trim=-0cm 0mm 0mm 0mm,scale=0.38]{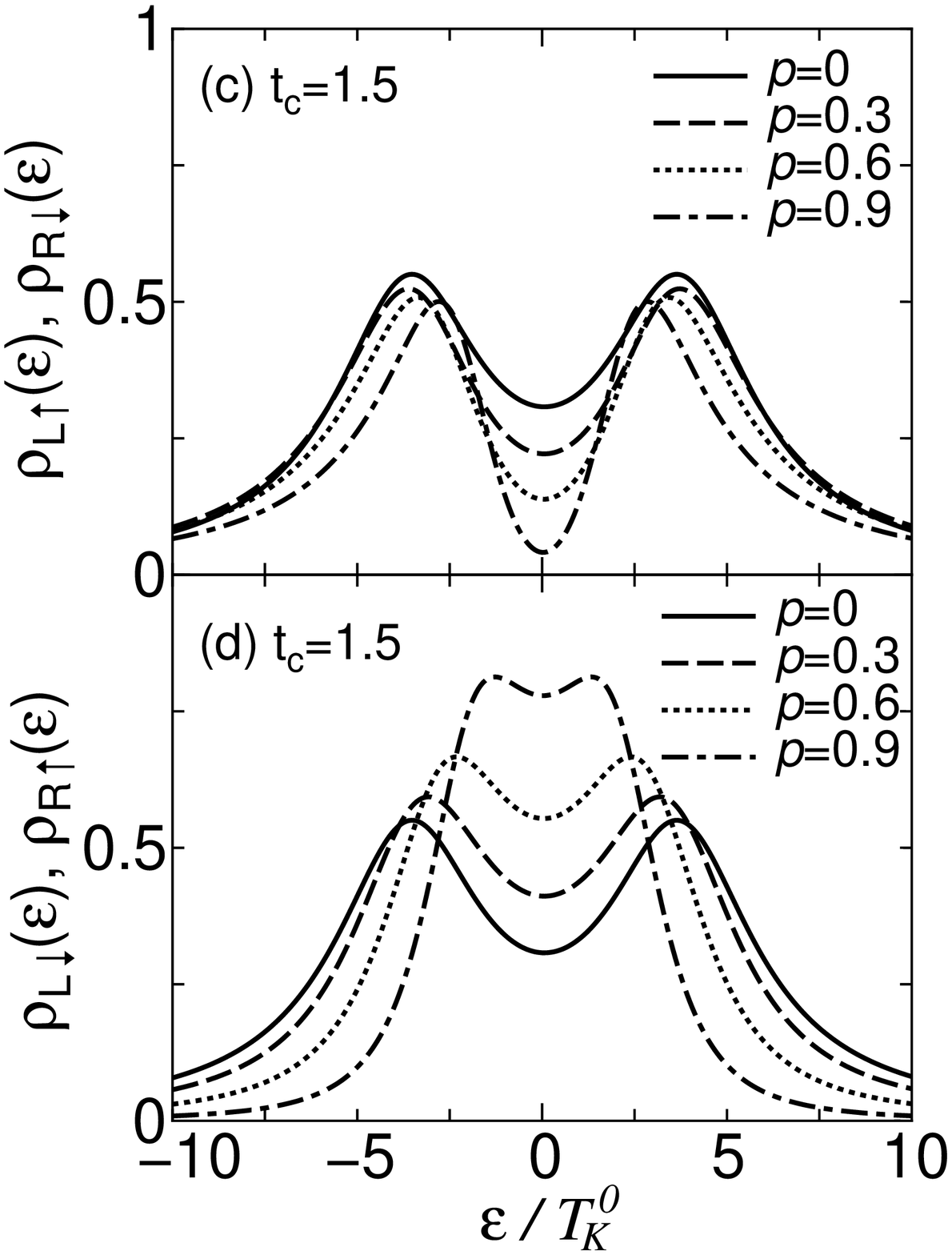}
\end{minipage}
\caption{
DOS of the left and right dots at  various $p$:
$t_c=0.3$ in (a) and (b);   $t_c=1.5$ in (c) and (d).
}
\label{dos15}
\end{figure}

Similarly to the P case, the characteristics in the conductance
can be interpreted in terms of the modified Kondo resonances.
In Fig. \ref{dos15} we show the local DOS.
According to  Eq.\eqref{resonance2}, we see that
  $\rho_{L\uparrow}(\varepsilon )$ ($\rho_{L\downarrow}(\varepsilon )$)
is identical to
$\rho_{R\downarrow}(\varepsilon )$ ($\rho_{R\uparrow}(\varepsilon )$).
We focus on the up-spin electrons for simplicity.
$\rho_{R\uparrow}(\varepsilon)$
and $\rho_{L\uparrow}(\varepsilon )$ are mixed with each other in 
the presence of the coupling $t_c$.
We note that  the right  (left) resonance 
gets narrower (broader) with the increase of $p$, so that 
 for $t_c=0.3$ there appears a dip structure 
in $\rho_{L\uparrow}(\varepsilon )$ around $\varepsilon=0$, 
while a single peak 
structure still remains in $\rho_{R\uparrow}(\varepsilon )$, 
as seen from 
Figs. \ref{dos15}(a) and (b). Recall that the conductance 
is roughly given by the 
product of $\rho_{R\uparrow}(\varepsilon )$ 
and $\rho_{L\uparrow}(\varepsilon )$
at the Fermi level.
Therefore, we naturally see that the combination  of the 
peak and dip structures in the DOS
 gives a maximum structure in the conductance for $t_c=0.3$.
On the other hand, for $t_c=1.5$, the development of the peak structure
in $\rho_{R\uparrow}(\varepsilon )$ is slow with the increase of
$p$. Thus the development of the
dip structure in $\rho_{L\uparrow}(\varepsilon )$ dominates
the slight change in  $\rho_{R\uparrow}(\varepsilon )$, leading 
to the monotonic decrease of the 
conductance as a function of $p$, as seen in Fig. \ref{APp-g}.
Note that at $p=1$, the current vanishes, in 
contrast to the P configuration, because either of 
the DOS of the right or left dot vanishes at the Fermi level.

\subsubsection{Nonequilibrium systems $(V\ne 0)$}

We finally show the results in the nonequilibrium 
case with  the AP configuration.
\begin{figure}[h]
\begin{minipage}{.80\linewidth}
\includegraphics[trim=-0cm 0mm 0cm 0mm,scale=0.4]{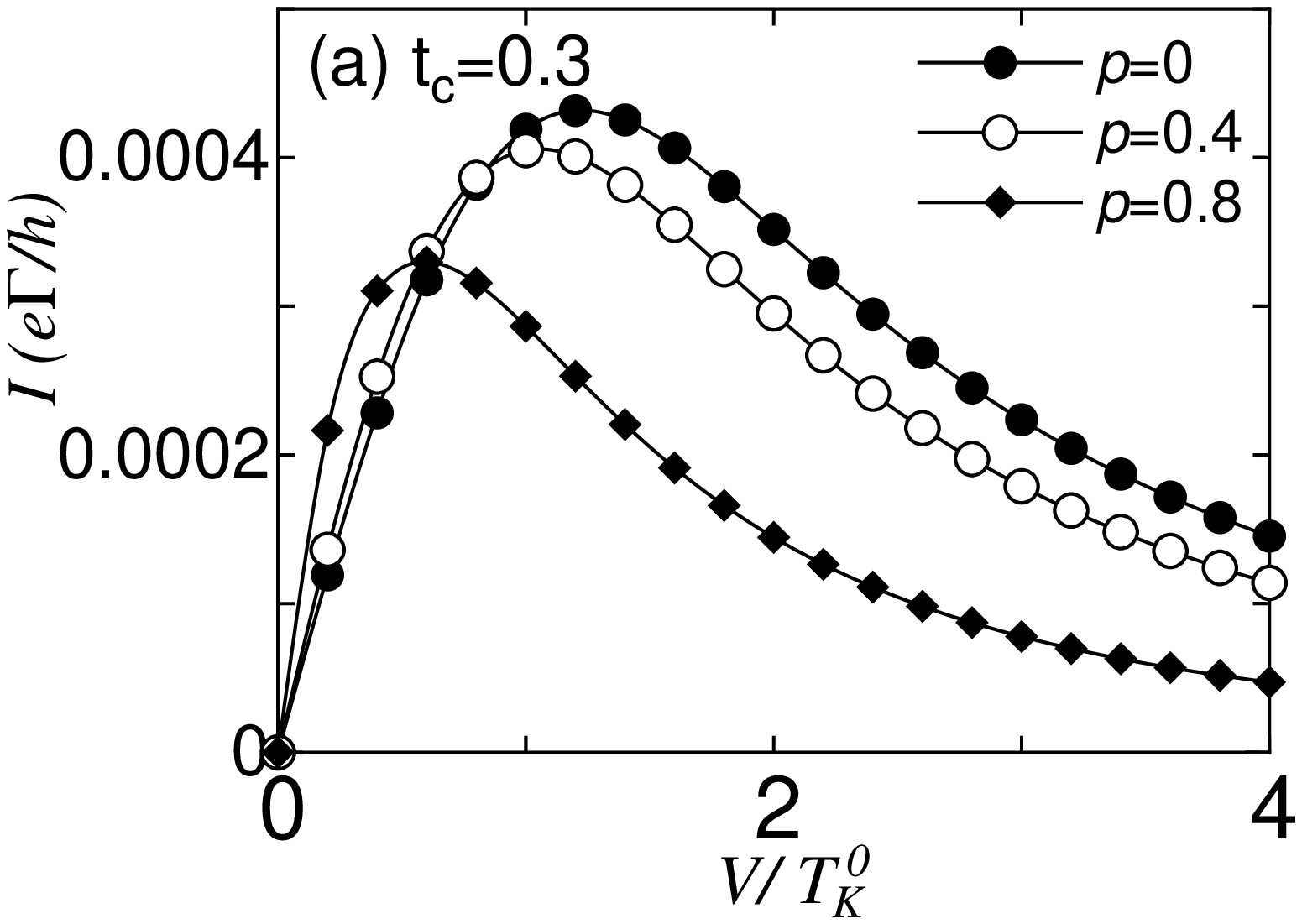}
\end{minipage}
\begin{minipage}{.80\linewidth}
\includegraphics[trim=-0.0cm 0mm 0mm 0mm,scale=0.4]{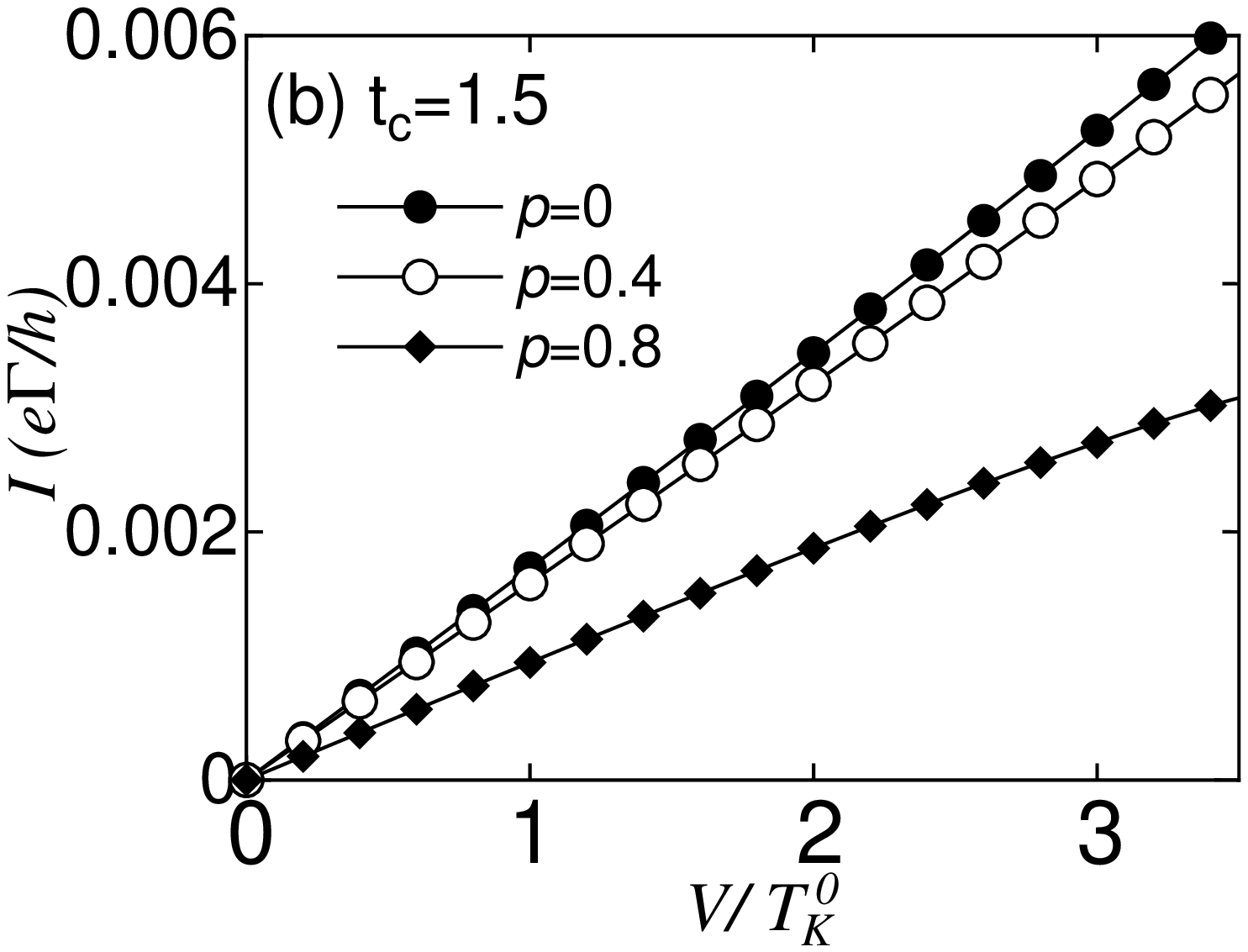}
\end{minipage}
\caption{
{\it I-V} characteristics for the AP configuration
 at $p=0, 0.4$ and $0.8$:
(a)$t_c=0.3$ and (b)$t_c=1.5$.
}
\label{APiv}
\end{figure}
\begin{figure}[h]
\begin{minipage}{.90\linewidth}
\includegraphics[trim=-0.4cm 0mm 0cm 0mm,scale=0.4]{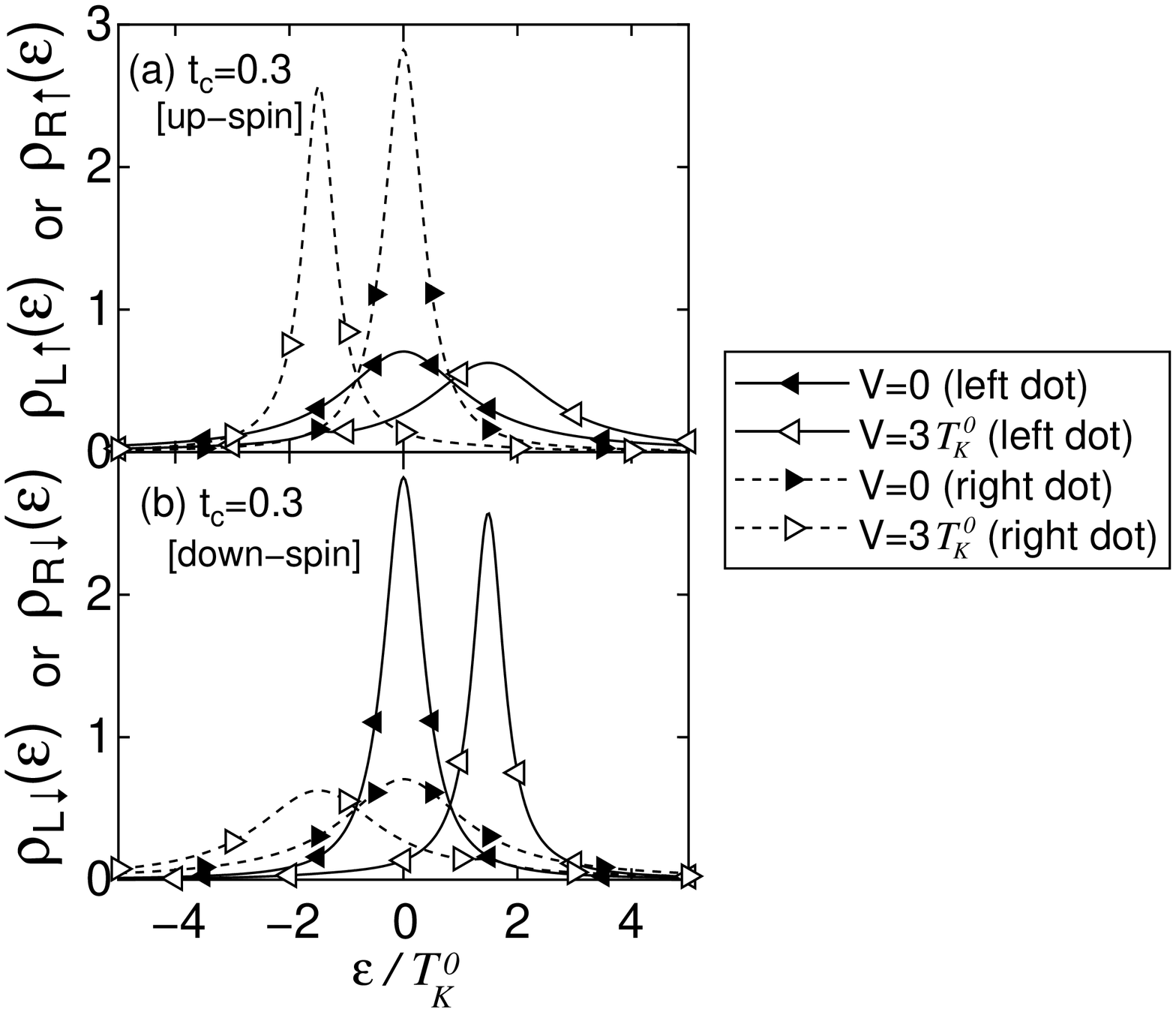}
\end{minipage}
\begin{minipage}{.90\linewidth}
\includegraphics[trim=-0cm 0mm 0mm 0cm,scale=0.4]{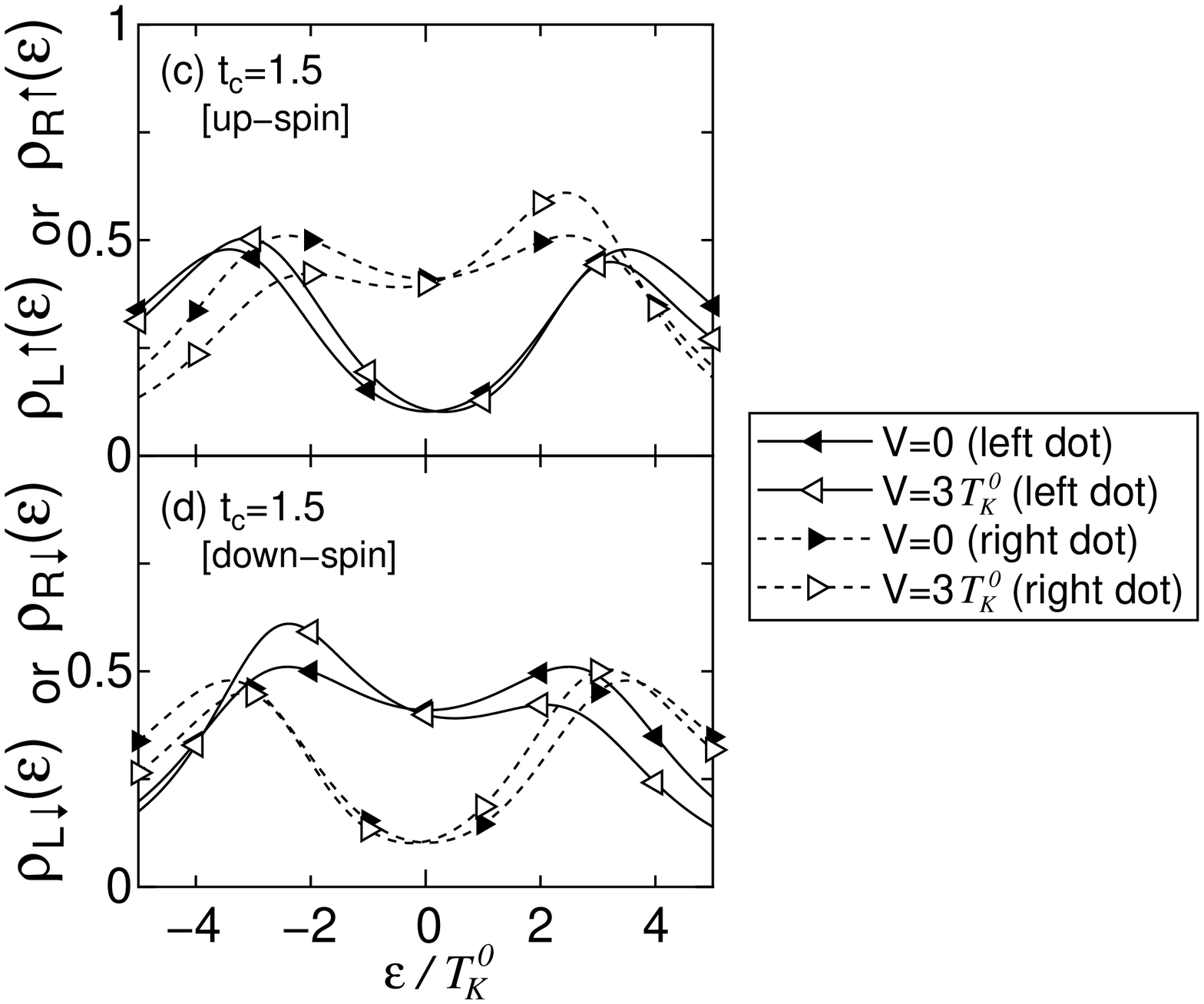}
\end{minipage}
\caption{
DOS of the left dot (solid line) and right dot (broken line) for the bias
voltage $V=0$ and $3T_K^0$ at $p=0.6$: (a) up- and (b) down-spin 
electrons in the case of $t_c=0.3$;
(c) up- and (d) down-spin electrons in the case of $t_c=1.5$.
}
\label{APdos}
\end{figure}
Figure \ref{APiv} shows the {\it I-V} characteristics of $t_c=0.3$ 
and $1.5$ for various choices of $p$.  We note again that the
$V$-dependence of the total current 
shows  similar behavior to that for the P
 configuration (Fig. \ref{Piv}) although the origin of the
characteristics is somewhat different, as mentioned above. 
Shown in Fig. \ref{APdos} are the Kondo resonances in nonequilibrium 
case at $p=0.6$.
When the bias voltage is applied, the peak position of 
$\rho_{L\sigma}(\varepsilon )$, $\rho_{R\sigma}(\varepsilon )$
$(\sigma =\,\uparrow$ or $\downarrow)$
 follows the change of the left or right chemical potential,
as shown in Figs. \ref{APdos}(a) and (b).
As mentioned above, the current for up- or down-spin electrons 
is controlled by the overlap of the Kondo resonances of
the left and right dots in the region of
$\varepsilon =-V/2$ to $\varepsilon =V/2$ in Fig. \ref{APdos}.
This gives rise to the maximum structure in the 
{\it I-V} characteristics for $t_c=0.3$.
 As seen in Figs. \ref{APdos}(c) and (d) for $t_c=1.5$,
the shape of  the Kondo resonances with splitting
is not altered so much up to a certain voltage, giving 
a linear conductance, as is the case for the P configuration.
For larger voltage, the system suffers from a sudden 
phase transition
accompanied by discontinuous drop in $I$ \cite{Lang}.

\section{Summary}

We have discussed  equilibrium and non-equilibrium transport 
properties of double quantum dots 
attached to the FM leads in series.
It has been found that, under the influence of the FM leads,
transport  shows the characteristic properties due to 
the modified Kondo resonances. We have clarified that the 
total conductance shows analogous $p$-dependence between P and
AP configurations of the leads, although the origin of such 
behavior is somewhat different from each other.

For the P configuration, at small
inter-dot coupling $t_c$, the conductance of down-spin (minority spin)
electrons dominates that of up-spin (majority spin) electrons,
giving rise to a peak structure in the $p$-dependent conductance.
On the other hand, for large coupling $t_c$, the conductance 
 exhibits monotonic  $p$-dependence, and is mainly contributed 
by up-spin electrons in  the large $p$ region.
For the AP configuration, we have observed similar $p$-dependence
in the total conductance, although
the contribution of the up- or down-spin electrons is always same
in this case.  

We have also studied   the
current $I$ at finite bias voltage, which again shows  distinct properties 
depending on $t_c$. For small inter-dot 
coupling, the Kondo resonances of
the left and right dots are shifted,
following  the change in the chemical potential. This gives rise to
 a characteristic peak structure in the $V$ dependence 
of the current.  On the other hand, for large inter-dot 
coupling, the shape of the modified Kondo resonances
is not changed so much up to large $V$, leading to
 the linear conductance in the 
wide $V$ region, beyond which the systems 
undergoes  a first-order
 transition
accompanied by the sudden decrease of the current.

Experimentally, the Kondo effect in double dot systems has 
 been observed recently\cite{Jeong,Chen}.  We thus 
expect that in the near future, it may become possible to
observe the Kondo effect in double dot systems coupled to 
the  FM leads, providing further interesting 
examples of correlation effects in the context 
of spin-dependent transport in nanoscale systems.

\begin{acknowledgments}
The work is partly supported by a Grant-in-Aid from the
 Ministry of Education, Culture,
Sports, Science and Technology of Japan.
\end{acknowledgments}


\begin{thebibliography}{99}

\bibitem{Gold} 
D. Goldhaber-Gordon, H. Shtrikman, D. Mahalu, D. Abusch-Magder, U. Meirav and M. A. Kanster: 
Nature \textbf{391} (1998) 156 ;
 D. Goldhaber-Golden, J. G$\ddot{\textrm{o}}$res, M. A. Kastner, H. Shtrikman, D. Mahalu and U. Meirav: 
Phys. Rev. Lett. \textbf{81} (1998) 5225.

\bibitem{Cronen} 
S. M. Cronenwett, T. H. Oosterkamp and L. P. Kouwenhoven:
Science \textbf{281} (1998) 540.

\bibitem{Ng} 
T. K. Ng and P. A. Lee: 
Phys. Rev. Lett. \textbf{61} (1988) 1768.

\bibitem{Glaz} 
L. I. Glazman and M. E. Raikh: 
JETP Lett. \textbf{47} (1988) 452.

\bibitem{Kawa} 
A. Kawabata: 
J. Phys. Soc. Jpn. \textbf{60} (1991) 3222.

\bibitem{MW} 
Y. Meir and N. S. Wingreen: 
Phys. Rev. Lett. \textbf{68} (1992) 2512; 
Y. Meir, N. S. Wingreen and P. A. Lee: 
Phys. Rev. Lett. \textbf{70} (1993) 2601;
A. -P. Jauho, N. S. Wingreen and Y. Meir:  
Phys. Rev. B \textbf{50} (1994) 5528.

\bibitem{Ogu} A. Oguri, H. Ishii and T. Saso: Phys. Rev. B \textbf{51} (1995) 4715.

\bibitem{Sakai} 
W. Izumida, O. Sakai and Y. Shimizu: 
J. Phys. Soc. Jpn. \textbf{67} (1998) 2444.

\bibitem{p2} 
B. Wang, J. Wang and H. Guo: 
J. Phys. Soc. Jpn. \textbf{70} (2001) 2645.

\bibitem{Zhang} 
P. Zhang, Q. -K. Xue, Y. Wang and X. C. Xie: 
Phys. Rev. Lett. \textbf{89} (2002) 286803.

\bibitem{p1} 
N. Sergueev, Q. -f. Sun, H. Guo, B. G. Wang and J. Wang: 
Phys. Rev. B \textbf{65} (2002) 165303.

\bibitem{JiMa} 
J. Ma, B. Dong and X. L. Lei: 
cond-mat/0212645.

\bibitem{Lopez} 
R. L$\acute{\textrm{o}}$pez and D. S$\acute{\textrm{a}}$nchez: 
Phys. Rev. Lett. \textbf{90} (2003) 116602. 

\bibitem{Koing} 
J. K$\ddot{\textrm{o}}$nig and J. Martinek: 
Phys. Rev. Lett. \textbf{90} (2003) 166602. 

\bibitem{Utumi} 
J. Martinek, Y. Utsumi, H. Imamura, J. Barna$\acute{\textrm{s}}$, S. Maekawa,
 J. K$\ddot{\textrm{o}}$nig and G. Sch$\ddot{\textrm{o}}$n: 
Phys. Rev. Lett. \textbf{91} (2003) 127203.

\bibitem{BinDon} 
B. Dong, H. L. Cui, S. Y. Liu and X. L. Lei: 
cond-mat/0302372.

\bibitem{Choi} 
M. -S. Choi, D. S$\acute{\textrm{a}}$nchez and R. L$\acute{\textrm{o}}$pez: 
Phys. Rev. Lett. \textbf{92} (2004) 056601. 

\bibitem{Julli} 
M. Julliere: 
Phys. Lett. \textbf{54A} (1975) 225.

\bibitem{Fert} M. N. Baibich, J. M. Broto, A. Fert, F. Nguyen Van Dau, F. Petroff, P. Eitenne, G. Creuzet, A. Friederich and J. Chazelas: 
Phys. Rev. Lett. \textbf{61} (1988) 2472.

\bibitem{Slon} 
J. C. Slonczewski:
Phys. Rev. B \textbf{39} (1989) 6995.

\bibitem{Prinz} 
G. A. Prinz:
Science \textbf{282} (1998) 1660.

\bibitem{Wolf} S. A. Wolf, D. D. Awschalom, R. A. Buhrman, J. M. Daughton, S. von Moln$\acute{\textrm{a}}$r, M. L. Roukes, A. Y. Chtchelkanova and D. M. Treger: Science \textbf{294} (2001) 1488.


\bibitem{Eto2}
T. Aono, M. Eto and K. Kawamura: 
J. Phys. Soc. Jpn. \textbf{67} (1998) 1860.

\bibitem{George}
A. Georges and Y. Meir:
Phys. Rev. Lett. \textbf{82} (1999) 3508.

\bibitem{Lang}
R. Aguado and D. C. Langreth:
Phys. Rev. Lett. \textbf{85} (2000) 1946.

\bibitem{Izumi}
W. Izumida and O. Sakai:
Phys. Rev. B \textbf{62} (2000) 10260.

\bibitem{Eto} 
T. Aono and M. Eto: 
Phys. Rev. B \textbf{63} (2001) 125327.

\bibitem{Rosa} 
R. L$\acute{\textrm{o}}$pez R. Aguado and G. Platero:
Phys. Rev. Lett. \textbf{89} (2002) 136802.


\bibitem{Jeong} 
H. Jeong, A. M. Chang and M. R. Melloch:
Science \textbf{293} (2001) 2221.

\bibitem{Chen} 
J. C. Chen, A. M. Chang and M. R. Melloch: 
Phys. Rev. Lett. \textbf{92} (2004) 176801.

\bibitem{Tae} 
T. -S. Kim and S. Hershfield: 
Phys. Rev. B \textbf{63} (2001) 245326.

\bibitem{Kikoin} 
K. Kikoin and Y. Avishai:
Phys. Rev. Lett. \textbf{86} (2001) 2090;
K. Kikoin and Y. Avishai: 
Phys. Rev. B \textbf{65} (2002) 115329.


\bibitem{Taka} 
Y. Takazawa, Y. Imai and N. Kawakami: 
J. Phys. Soc. Jpn. \textbf{71} (2002) 2234.
\bibitem{AC} 
A. C. Hewson: 
\textit{The Kondo Problem to Heavy Fermions} (Cambridge University Press, Cambridge, 1993).

\bibitem{Cole} 
P. Coleman: 
Phys. Rev. B \textbf{29} (1984) 3035.
%
\bibitem{Keldysh} H. Haug and A. -P. Jauho: \textit{Quantum Kinetics in Transport and Optics of Semi-conductors},
 edited by M. Cardona et al. (Springer-Verlag, 1998).


\end{thebibliography}
\end{document}